\documentclass[12pt,oneside]{article}

\begin{document}

\thispagestyle{empty} \setcounter{page}{0} \renewcommand{\theequation}{%
\thesection.\arabic{equation}}

{\hfill{CECS-PHY-02/01}}

\vspace{2cm}

\begin{center}
\textbf{GRAVITATIONAL THERMODYNAMICS OF SCHWARZSCHILD DE SITTER SPACE}

\vspace{1.4cm}

CLAUDIO TEITELBOIM

\vspace{.2cm}

\emph{Centro de Estudios Cient\'{i}ficos } \\[0pt]
\emph{Valdivia - Chile } \\[0pt]
\end{center}

\vspace{-.1cm}

\centerline{{\tt teitel@cecs.cl}}

\vspace{1cm}

\centerline{ABSTRACT}

\vspace{- 4 mm}

\begin{quote}
{\small The Euclidean Schwarzschild-de Sitter geometry may be considered as
an extremum of two different action principles. If the thermodynamical
parameters are held fixed at the cosmological horizon, one deals with the
gravitational thermodynamical effects of the black hole but ignores those of
the cosmological horizon. Conversely, if the macroscopical variables are
held fixed at the black hole horizon, it is only the cosmological horizon
thermodynamics which is dealt with. Both cases are analyzed. In particular,
the internal energy }$U$ {\small is calculated in the semiclassical
approximation as a function of the mass parameter }$m$ {\small of
Schwarzschild de Sitter space. In the first case one finds }$U=+m${\small ,
while in the second one gets }$U=-m${\small . This suggests that de Sitter
space is thermodynamically unstable under black hole formation. }
\end{quote}

\baselineskip18pt

\vskip 2cm

{\small To be published in the proceedings of the Francqui Colloquium ``Strings and Gravity: 
Tying the Forces Together", Brussels, October 19 - 21, 2001. Edited by Marc Henneaux and 
Alexander Sevrin.}

\newpage

\setcounter{equation}{0}

\section{Introduction}

The observational evidence for a positive cosmological constant has led to
renewed interest in the dynamics of de Sitter space. It is natural in this
context to analyze the thermodynamics of de Sitter space in the presence of
a black hole. It has been known for a long time\cite{GH} that
if one tries to use the Euclidean Schwarzschild-de Sitter solution to
provide thermodynamical information, one finds that the time periods
required to avoid a conical singularity at both, the cosmological and black
hole horizons, do not match. This is physically interpreted as indicating
that the two horizons are not in thermal equilibrium and that, for example,
they both emit Hawking radiation at the corresponding temperatures.
An observer somewhere in space would then
receive a beam of radiation coming from the black hole and,
at the same time, isotropic radiation coming from the
cosmological horizon.

>From the point of view of the action principle, the fact that the
Schwarzschild-de Sitter solution cannot be made to have no conical
singularity means that the empty space field equations are not satisfied
everywhere. If one arranges the period of the time variable so as to have no
conical singularity at the cosmological horizon, the field equations will be
satisfied there but will not be satisfied at the black hole horizon.
Conversely, if the role of the horizons is interchanged, the field equations
will not be satisfied at the cosmological horizon.

It is the main purpose of this article to point out that this apparent
difficulty is rather a virtue and it was to be expected from the point of
view of the action principle and thermodynamics. Indeed, if one uses an
extremum of the action to evaluate the path integral in the semiclassical
approximation, one needs to hold fixed those variables which will become the
argument of the partition function once it is evaluated. By the very meaning
of ''holding fixed'', those variables are not varied in the action
principle. Thus, for example, for a black hole in asymptotically flat space,
one may hold fixed the \textrm{1/r} part of the components of the metric
which are determined by the mass. Then, one is dealing with the
microcanonical ensemble, where the partition function depends on the total
energy. It would be wrong to demand that the partition function thus
obtained should have an extremum with respect to variations in the \textrm{%
1/r} piece of the spatial metric, because then one would obtain a particular
value for the mass, i. e., $\mathrm{m=0}$, and thus would not have enough information
to develop the thermodynamics of the system.

For the case at hand, there is no notion of spacelike infinity, but the
problem itself indicates what to do. One may fix appropriate components of
the metric at either the cosmological or black hole horizons. If one chooses
the cosmological horizon as the place where the variables are held fixed,
there will be no field equations to satisfy at that point.
Then the cosmological horizon will be the analog of
spatial infinity in the asymptotically flat case, where the "observer" sits
(or, more precisely, the analog of a very large sphere whose radius is
eventually allowed to grow without limit). The problem one is solving then
will be the thermodynamics of a black hole contained in a space of a given
cosmological radius ("box", "boundary"). Conversely, if the variables are fixed
at the black hole horizon, it is then that horizon which acts as the
boundary. One would then be discussing the thermodynamics of a cosmological
horizon with the black hole acting as the boundary. Changes in the black
hole variables would then not be subject  to dynamics but rather would
correspond to changes that the ``external observer" decides to make in the
environment.

This discussion shows that one should be able to use the Schwarzschild-de
Sitter solution as a true extremum of two different action principles which
correspond to two different thermodynamical problems. One problem is the
thermodynamics of a black hole horizon with a cosmological boundary. The
other is the thermodynamics of a cosmological horizon with a black hole
boundary. It turns out, as we shall see below, that the physical
properties of the two systems have some striking differences.

\section{Action Principles for the Euclidean Schwarzschild-de Sitter Metric}

\setcounter{equation}{0}

The (Euclidean) Schwarzschild-de Sitter metric may be written as
\begin{equation}
ds^{2}=f^{2}dt^{2}+f^{-2}dr^{2}+r^{2}(d\theta ^{2}+\sin ^{2}d\varphi ^{2}).
\label{arc}
\end{equation}
Here $\theta $ and $\varphi $ are the usual coordinates on the two-sphere,
the time variable $t$ is periodic, with a period that will be discussed
below, and the radial variable $r$ runs over a range that will also be
discussed in what follows. The function $f^{2}$ appearing in the line
element is given by
\begin{equation}
f^{2}=1-\frac{2m}{r}-\frac{r^{2}}{l^{2}}.  \label{f2}
\end{equation}
It depends on two parameters the mass $m$, and the cosmological radius $l$,
which is related to the cosmological constant $\Lambda $ by $l=\left( \frac{3%
}{\Lambda }\right) ^{1/2}$, and it has two roots, $r_{+}(m,l)$ and $%
r_{++}(m,l)$,

\begin{equation}
f^{2}(r_{+}(m,l))=f^{2}(r_{++}(m,l))=0.  \label{roots}
\end{equation}
These two solutions exist and are different if and only if
\begin{equation}
27\frac{m^{2}}{l^{2}}< 1
\label{inequality}
\end{equation}

The smaller root $r_{+}(m,l)$ will be called the black hole horizon radius,
and the larger solution $r_{++}(m,l)$ will be called the cosmological
horizon radius. Thus, $r_{+}<r_{++}$. Because of (\ref{roots}), $r_{+}$ and $r_{++}$
are single points, rather than circles in $r-t$ space.

As explained in the introduction, we will be interested in two different
cases, one in which $r_{++}$ is treated as a boundary and the other in which
$r_{+}$ is a boundary. In each case the corresponding point will be removed
from the manifold and thus $r-t$ space will be a disc rather than a
two-sphere. When  $r_{++}$ is treated as a boundary we will be including the
thermodynamics of the black hole horizon. This is because  $r_{+}$ will be
varied then, which stems from  the fact that one is integrating over
black hole horizon configurations in the partition function. For this
reason, we will call this case the black hole case. Thus we have
\[
r_{+}\leq r<r_{++} \ \ \ \ \ \ \ \mbox{ (black hole
case).}
\]
and, similarly

\[
r_{+}<r\leq r_{++}\ \ \ \ \ \ \ \ \ \mbox{ (cosmological case)}
\]

In the black hole case one must demand that the empty space field equations
should be satisfied at $r_{+}$. This is because  $r_{+}$ is included in the
manifold. In order for this requirement to follow from the action principle,
one must take the action to be equal to the sum of the Hamiltonian action
and one quarter of the black hole horizon area \cite{Gauss-Bonnet}

\begin{equation}
I_{\mbox{\tiny Black hole}}=\frac{1}{4}4\pi r_{+}^{2}+I_{\mbox{\tiny Hamiltonian}},
\label{Action}
\end{equation}
where
\begin{equation}
I_{\mbox{\tiny Hamiltonian}}=\int (N^{\bot }\mathcal{H}_{\bot }+N^{i}\mathcal{H}%
_{i})dt\mbox{ }d^{3}x.  \label{Ham Action}
\end{equation}
(We use the convention that one functionally integrates $\exp(+I_{\mbox{\tiny Euclidean}})$).

When one demands that the action (\ref{Action}) should have an extremum
under variation of $r_{+}$, the variation of the area term combined with a
contribution arising from an integration by parts in the variation of (\ref
{Ham Action}), yields the value

\begin{equation}
\beta _{+}=\frac{4\pi }{(f^{2})^{\prime }(r_{+})},  \label{Beta}
\end{equation}
for the period of the Euclidean time variable, which is equal to the inverse
of the temperature of the black hole.

The value (\ref{Beta}) will cause a conical singularity at $r_{++}$. But
this is not a problem since $r_{++}$ is not on the manifold and is not
varied. Thus, the action (\ref{Action}) has a true extremum even though
there is a conical singularity at $r_{++}$. This means, in particular, that,
in this formulation, no thermodynamical properties are associated with the
cosmological horizon and thus $\beta _{+}$ is not its Hawking temperature.

Conversely, for the cosmological case one has
\begin{equation}
I_{\mbox{\tiny cosmological}}=\frac{1}{4}4\pi r_{++}^{2}+I_{\mbox{\tiny Hamiltonian}},
\label{Actionplusplus}
\end{equation}
where $I_{\mbox{\tiny Hamiltonian}}$ is again given by (\ref{Ham Action}).

This time one finds

\begin{equation}
\beta _{++}=-\frac{4\pi }{(f^{2})^{\prime }(r_{++})}.  \label{Betaminus}
\end{equation}
The minus sign on the right hand side of (\ref{Betaminus}) arises directly
from the variation of (\ref{Actionplusplus}) and it is quite reasonable
because $(f^{2})^{\prime}(r_{++})$ is negative.
This sign will have important consequences below.

\section{Thermodynamic Functions}
In the semiclassical approximation, the value of $I_{\mbox{\tiny Hamiltonian}}$ is equal to zero. This is because
the constraints ${\cal H}_{\perp}={\cal H}_{i}=0$, which are four of the ten ``bulk" field equations hold
and, furthermore, $\dot{g}_{ij}$ also vanishes because the Shwarzschild--de Sitter metric is stationary
(time independent). Thus, in either the black hole or cosmological cases, the value of the Euclidean action
on shell reduces to just one fourth of the corresponding horizon area. Since in either case  the quantity held fixed
maybe taken to be the parameter $m$, which, as it will be seen below, will be a function of the internal energy,
one is dealing with the microcanonical ensemble. Thus one fourth of the
area is the entropy,
\begin{eqnarray}
S_{+} &=& \pi r_{+}^2(m) \ \ ,   \\
S_{++} &=& \pi r_{++}^2(m) \ \ .
\label{entropy}
\end{eqnarray}

 However,  the relationship between the parameter $m$ and the internal energy $U$ offers
a bit of a surprise. To see this it is best to begin with the first law of thermodynamics,
\begin{equation}
\beta dS = dU \ \ .
\label{firstlaw}
\end{equation}
We have for the variation of the entropy
\begin{equation}
\label{ds}
dS = 2\pi r_{\mbox{\tiny H}} dr_{\mbox{\tiny H}}  \ ,
\end{equation}
where $r_{\mbox{\tiny H}}$ stands for $r_{+}$ or $r_{++}$ depending on whether one is dealing with the
black hole or the cosmological cases respectively. We obtain $dr_{\mbox{\tiny H}}$ by differentiating (\ref{roots}), which gives
\begin{equation}
(f^2)'(r_{\mbox{\tiny H}})dr_{\mbox{\tiny H}}  + \frac{\partial f^2}{\partial m }(r_{\mbox{\tiny H}}) dm = 0 \ ,
\end{equation}
which, recalling (\ref{f2}), yields in turn,
\begin{equation}
dr_{\mbox{\tiny H}} = \frac{1}{(f^2)'(r_{\mbox{\tiny H}})} \frac{2}{r_{\mbox{\tiny H}}} dm \ .
\label{drh}
\end{equation}
If we insert back (\ref{drh}) in (\ref{ds}) we find
\begin{equation}
dS = \frac{4\pi}{(f^2)'(r_{\mbox{\tiny H}})} dm \ .
\label{DS}
\end{equation}
Next,  recalling  (\ref{Beta}) and (\ref{Betaminus}) we obtain
\begin{equation}
dS_{+} = \beta_+ dm
\label{fl1}
\end{equation}
and
\begin{equation}
dS_{++} = -\beta_{++} dm \ .
\label{fl2}
\end{equation}
Finally, comparing (\ref{fl1}) and (\ref{fl2}) with the first law of thermodynamics (\ref{firstlaw}) we conclude
that, up to an irrelevant constant of integration, the internal energies of the black hole and cosmological
horizons are given by
\begin{equation}
U_+= + m
\end{equation}
and
\begin{equation}
U_{++} = -m \ .
\end{equation}
Thus, we find a situation analogous to that of electric
charge on a two--sphere. In that case, if a charge $q$ is placed at the
North Pole, an opposite charge, $-q$, appears at the South Pole. This is due
to the fact that the lines of force which diverge from the North Pole converge
onto the South Pole. The same phenomenon occurs here for the energy in $r-t$ space.

[A negative energy has also been associated with Schwarzschild--de Sitter space in \cite{BBM}.]

One may also calculate, in both cases, the Helmholtz free energy $F$ given by
\begin{equation}
-\beta F = S - \beta U  \ ,
\end{equation}
which amounts to add to the corresponding Euclidean action  the  term
$-\beta_+ m$ for the black hole case or $+\beta_{++}m$ for the cosmological case. It is important to emphasize, that in the
black hole case, $m$ is to be understood as a surface term on the cosmological horizon, that is, a function of $r_{++}$
obtained from $f^2(r_{++})=0$, with $r_{++}$ being the largest of the two roots of $f^2$. This is the analog of the mass
being a surface integral at infinity in the asymptotically flat case.
Conversely, in the cosmological case, $m$ is to be understood as a boundary term at $r_+$.
In either case, the surface term implements the Legendre transformation needed to pass from the microcanonical ensemble to
the canonical one, and it amounts to keep fixed  the temperature instead of
the internal energy at $r_+$ or at $r_{++}$ respectively .

Finally we point out that the specific heat $C$ of both systems, the black hole horizon  with the
cosmological boundary, or the cosmological horizon with the black hole boundary, is negative. This may be seen as follows.
By definition,
\begin{equation}
C^{-1}=\frac{d\beta^{-1}}{dU}= \frac{1}{4\pi} \frac{d}{dU} (f^2)'(r_{\mbox{\tiny H}}) \ ,
\label{c}
\end{equation}
but one may verify by a simple, straightforward calculation that
\begin{equation}
\frac{d}{d U} (f^2)'(r_{\mbox{\tiny H}}) = - \left(\frac{1}{r_{\mbox{\tiny H}}^2} + \frac{3}{l^2}\right)
\frac{d r_{\mbox{\tiny H}}}{d U} \ .
\label{df2du}
\end{equation}
However, it follows from (\ref{drh}) that an increase in the mass $m$ brings the two horizons closer together, thus one has
\begin{equation}
\frac{d r_+}{d m} > 0 \ ,  \ \ \ \  \frac{d r_{++}}{d m} < 0 \ \ .
\end{equation}
Therefore
\begin{equation}
\frac{dr_{\mbox{\tiny H}}}{dU} > 0 \ ,
\end{equation}
which, when combined with (\ref{df2du}) shows that the specific heat, $C$, given by  (\ref{c}), is negative
for both systems.

\section{Conclusion}

In the previous discussion we have pointed out that the Euclidean Schwarzschild--de Sitter line element may be
used to define two different, idealized physical systems. One of them is a black hole horizon enclosed in a cosmological
boundary. The other  is a cosmological horizon enclosed in a black hole boundary.
The geometrical structures of both systems are in close parallel  and so are their thermodynamics, but they also have striking
differences.

In actual physical circumstances, these two systems are not isolated, but rather they are coupled through the common parameter $m$,
since the (Lorentzian signature) solution includes both of them at once, away from thermal equilibrium. The question therefore arises as
to whether the two systems ever reach thermodynamical equilibrium, and, if so, what is the equilibrium configuration.  Of course, to
reach thermodynamical equilibrium one needs a process, for example  emission by the two horizons of membranes, {\em i.e.}, domain walls
with $3$--dimensional worldtubes, or simply Hawking radiation of particles.

The facts established in this paper suggest that the cosmological horizon would tend to lower its internal energy, thus {\em increasing}
the mass parameter $m$,and thus bringing closer to each other the two horizons $r_+$ and $r_{++}$. On the other hand, the black hole
horizon would tend to do the opposite. In particular, if initially the mass parameter is zero, so that one has pure the Sitter space, on would
expect the space to decay into a black hole formed out of the radiation emitted by the cosmological horizon. Once the black hole is formed,
the competition between both horizons will start. Two possible outcomes suggest themselves: ($i$) the cosmological horizon wins and the end
point is thermal equilibrium with the two horizons coalescing (the Nariai solution, which saturates (\ref{inequality})) or ($ii$) one could also speculate
on the possibility that  thermal equilibrium is never reached and one has transfer of energy to and fro that never ends. It would seem fair to say
that the answer to this question is not known at present.

\section{Acknowledgments}

The author is deeply indebted to Andr\'es Gomberoff, Marc Henneaux and Frank Wilczek for
many discussions.
This work  was funded by an institutional grant to CECS of the Millennium Science Initiative, Chile,
and also benefits from the generous support to CECS by Empresas CMPC.
 Partial support under FONDECYT grant 1010446 is also acknowledged.
The author wishes to thank the Francqui Foundation and the organizers of the meeting for their wonderful
hospitality.  The content of the present note developed out of the report presented at the Francqui Colloquium
and was first presented in the form reported here at the conference Quantum Gravity in the Southern Cone, held in Valdivia in January 2002.  The author also thanks
the organizers of that meeting for the opportunity to take part in it.

\end{document}